\documentclass{icrc2009}

\usepackage{graphicx}   % for including figures
\usepackage[caption=false]{caption}    % for captions
\usepackage[font=footnotesize]{subfig} % subfig.sty for a double column floating figure using two subfigures
\usepackage{fixltx2e}
\usepackage{url}

\newcommand{\shorttitle}[1]%
{\markboth{Proceedings of the 31\MakeLowercase{$^{st}$} ICRC, {\L}\'{o}d\'{z} 2009}{#1} }
\newcommand{\etal}{\MakeLowercase{\textit{et al. }}} % "et al."

\usepackage{amsmath,amssymb}
\usepackage{xspace}
\usepackage{psfrag}
\usepackage{lineno}
\newcommand{\xmax}{\ensuremath{X_\mathrm{max}}}
\newcommand{\gsm}{g/cm${}^2$}
\newcommand{\smilla}{\ensuremath{S_{1000}^\mathrm{tot}}}
\newcommand{\gss}{0.6}
\newcommand{\gs}{0.5}
\newcommand{\gsize}{0.39}

\psfrag{Xmax}[r][r][\gss]{\xmax, \gsm}
\psfrag{Stt[3][3]}[r][r][\gss]{$S_{1000}^\mathrm{tot}$, VEM}
\psfrag{Density}[r][r][\gss]{$D_{1000}$, m${}^{-2}$}

\psfrag{XmaxS1000_QGS}[c][c][\gs]{F(\xmax,\smilla)}
\psfrag{XmaxD1000_QGS}[c][c][\gs]{F(\xmax,$D_{1000}$)}
\begin{document} 

\title{New approach to primary mass composition analysis with
  simultaneous use of ground and fluorescence detectors data}

\author{\IEEEauthorblockN{A.~Yushkov\IEEEauthorrefmark{1},
                          M.~Ambrosio\IEEEauthorrefmark{1},
                          C.~Aramo\IEEEauthorrefmark{1},
                          F.~Guarino\IEEEauthorrefmark{1}\IEEEauthorrefmark{2},
                          D.~D'Urso\IEEEauthorrefmark{1},
                          L.~Valore\IEEEauthorrefmark{1}}
  \\
\IEEEauthorblockA{\IEEEauthorrefmark{1}INFN Sezione di Napoli, via Cintia, Napoli, Italia 80125}
\IEEEauthorblockA{\IEEEauthorrefmark{2}Universit\`{a} di Napoli ``Federico~II'', via Cintia, Napoli, Italia 80125}}

\shorttitle{A.~Yushkov \etal New approach to primary mass composition analysis}
\maketitle

\begin{abstract}
We study the possibility to reconstruct primary mass composition with
the use of combinations of basic shower characteristics, measured in
hybrid experiments, such as depth of shower maximum from fluorescence
side and signal in water Cherenkov tanks or in plastic scintillators
from the ground side. To optimize discrimination performance of shower
observables combinations we apply Fisher's discriminant analysis and
give statistical estimates of separation of the obtained distributions
on Fisher variables for proton and iron primaries. At the final stage
we apply Multiparametric Topological Analysis to these distributions
to extract composition from prepared mixtures with known fractions of
showers from different primary particles. It is shown, that due to
high sensitivity of water tanks to muons, combination of signal in
them with $\mathbf\xmax$ looks especially promising for mass
composition analysis, provided the energy is determined from
longitudinal shower profile.

\iffalse
The study is performed for 6000 showers of proton, oxygen and iron
primaries at 10~EeV and $\mathbf{37^\circ}$ zenith angle, generated
with CORSIKA QGSJET~01/Gheisha and QGSJET~II/Fluka.
\fi
\end{abstract}


\section*{Introduction}
The experimental information on UHECR mass composition coming from
different experiments and from different mass reconstruction
techniques is quite
contradicting~\cite{giller_jpg2007,gelmini_hecr2009}. One of the main
difficulties is that at these energies mass composition and hadronic
interactions properties are both unknown and are deeply entangled. The
necessary condition for the solution of this problem is the
reconciliation of the results on mass composition obtained from
different types of ground and fluorescence data between themselves and
with astrophysical predictions on the origin of the anisotropy,
`ankle' and GZK cut-off. Hybrid experiments, like Pierre Auger
Observatory~\cite{PAO_proto_NIMA2004} or Telescope Array
(TA)~\cite{TA_ISVHECRI2006}, are perfectly suitable for this purpose,
since they provide the opportunity to use combinations of extensive
air shower (EAS) parameters to achieve the best possible mass
resolution. The key role in this analysis can be played by the muon
shower content~---~the most problematic for hadronic
models~\cite{engel_icrc2007} and the best mass sensitive EAS
parameter. The upgrade of Auger with AMIGA scintillator counters
array~\cite{etche_icrc30_amiga} is aimed right at the muon content
measurement, but it is easy to show (see Section~\ref{sec:scan}) that
already the total signal in the Auger water tanks preserves the
difference between primaries in number of muons and can be profitable
for primary mass reconstruction, provided the energy is independently
determined from the longitudinal shower profile. In the case of TA,
which will be in grade to measure only charged particles density,
ground data alone will be weekly sensitive to primary particle mass
and idea of the use of EAS observables combinations becomes
indispensable. Using Auger and TA as examples, in this paper we put
forward a strategy allowing to reconstruct primary mass composition
from combinations of the fluorescence and ground data keeping in mind
the limitations on the affordable simulation statistics of the UHECR
showers.

\section{General notes on the choice of mass discrimination parameters}
\label{sec:scan}

In the following we consider cases of Auger and TA to estimate the
expected performance of the proposed mass reconstruction technique. We
assume that primary energy can be estimated from the longitudinal
shower profile and hence is practically primary mass
independent. Briefly speaking, to enhance primary mass resolution of
traditional \xmax\ parameter we suggest to use it in linear
combinations with other basic shower parameters, such as signal in
water tanks or particle density for Auger and TA correspondingly.

The data set used for the analysis was generated with
COR\-SI\-KA~6.204~\cite{corsika}
(QGSJET~01~\cite{qgsjet}/Gheisha~\cite{gheisha}) and COR\-SI\-KA~6.735
(QGSJET~II~\cite{qgsjetii}/Flu\-ka2008.3~\cite{fluka1}) packages and
contains 1000 showers for every primary (p, O, Fe) and interaction
model at 10~EeV and $37^\circ$ zenith angle. All longitudinal showers
characteristics and charged particles density were taken directly from
CORSIKA output files. The calculations of the expected signal in Auger
water tanks was performed according to the procedure described
in~\cite{billoir_sampl_2008,ave_munum_2007} with the use of the same
GEANT~4 lookup tables as in~\cite{ave_munum_2007}. In case of TA we
use charged particles density at 1000~m from the axis ($D_{1000}$) as
an example, for densities at another distances the consideration line
would be the same. Finally let us note, that qualitatively results for
both combinations of interaction models used in the study are very
similar, so below we will mostly discuss only results for
QGSJET~01/Gheisha, which provides some worse discrimination
performance. To characterize the separation of distributions we will
use the merit factor
$
\mathrm{MF}=|\bar x_\mathrm{Fe}-
  \bar  x_\mathrm{p}|/\sqrt{\sigma^2_\mathrm{Fe}+\sigma^2_\mathrm{p}},
$
where $\bar x_\mathrm{p,\,Fe}$ and $\sigma_\mathrm{p,\,Fe}$ are
distributions means and standard deviations correspondingly. 

\begin{figure}
\centering\includegraphics[width=\gsize\textwidth]{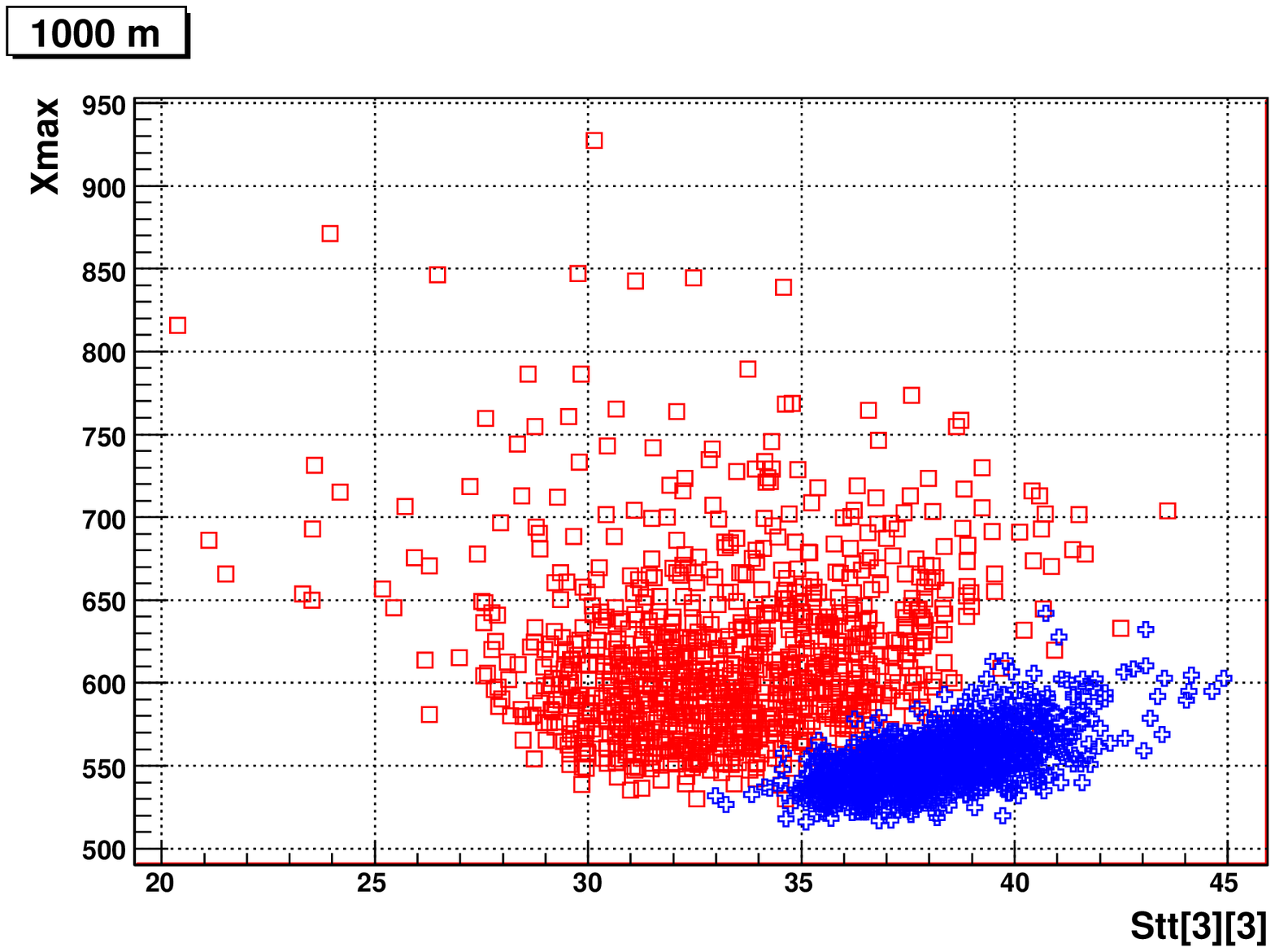}
\centering\includegraphics[width=\gsize\textwidth]{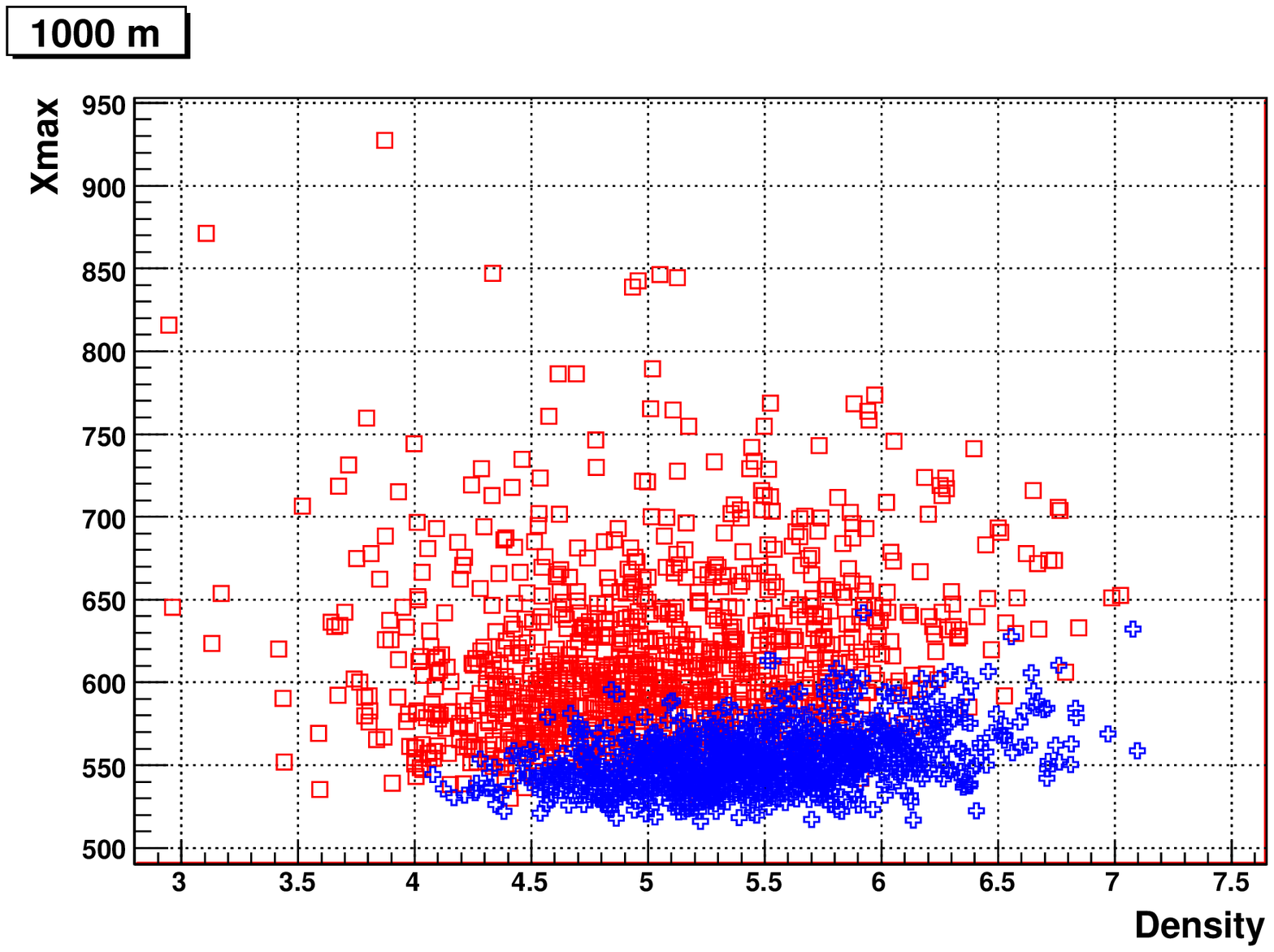}
\caption{Depth of shower maximum versus total ground plane signal and
  particle density for proton (red squares) and iron (blue crosses)
  showers at 1000 meters from the axis for QGSJET~01 model.}
\label{fig:sigxmax}
\end{figure}

In Fig.~\ref{fig:sigxmax} we present scatter plots of total ground
plane signal and charged particles density vs \xmax\ at 1000~m from
the shower axis. Good separation of iron and proton showers in
(\smilla,\,\xmax) plot is both due to discrimination power of
\xmax\ and to noticeable difference $\sim13$\% in the average total
signals. In absolute units this difference is the same as the
difference between muon signals, but the separation of the total
signals (MF=1.4) is surely worse that of the muon ones (MF=2.5) due to
the smearing effect of the electromagnetic component. Despite of
change with the distance of the (electromagnetic/muon) signal ratio, a
good separation of protons and iron nuclei is kept in a wide range
from 700 to 1500 meters, since high sensitivity of Cherenkov tanks to
the muon component results in different shifts of p and Fe populations
along signal axis in \xmax\ vs total signal scatter plots. On the
other hand, as expected, the separation between primaries in
($D_{1000}$,\,\xmax) plot is mostly due to discrimination power of
\xmax, since distributions on charged particles density for protons
and iron nuclei largely overlap (MF=0.5). In this case one can think
of searching for another discrimination parameters combinations, but
as it will be shown below, ($D_{1000}$,\,\xmax) pair provides one of
the best (among possible within TA conditions) discrimination
resolutions.

%\clearpage

\section{Fisher's discriminant analysis}
\label{sec:fisher}
The problem of primary particle mass discrimination with the use of
combination of two or more shower characteristics falls in the class
of standard tasks of statistical pattern classification analysis (see
e.g.~\cite{webb,kuncheva,TMVA}) and one of its methods -- linear
discriminant analysis -- was recently applied to study the
classification capability of longitudinal profile distribution
parameters~\cite{souza_LDA_07}. Using the Toolkit for Multivariate
Data Analysis (TMVA)~\cite{TMVA} here we perform a similar study for
combinations of different fluorescence and ground data.

As it was already discussed, p and Fe populations are well separated
in both examples in Fig.~\ref{fig:sigxmax}, and what's more, they can
be separated with high accuracy even by a straight line. Hence, it is
opportune to apply in this case just linear discriminant analysis and
namely Fisher's method. In this approach one seeks the direction along
which two classes will be separated the best, i.e. one looks for the
direction in parameters hyperspace, after projection on which the
ratio of the distance between distributions means to the sum of their
squared variations will be maximized. The evident advantage of this
approach is possibility to use any number of parameters avoiding
``dimensionality curse'' (thus reducing necessary simulations
statistics) and to apply easily any further classification tools to
the resulting one-dimensional distributions. In addition to Fisher's
discriminant the performance of rectangular cut optimization,
projective likelihood estimator and function discriminant analysis
with quadratic and cubic functions~\cite{TMVA} were checked and it was
found that none of them outperforms Fisher's approach.

To find the direction, along which the primaries will be separated in
the optimal way Fisher's algorithm requires minimum training: already
5--10 events of every primary type can be enough to achieve the same
results as in the case of the use of several hundreds events. After
application of Fisher's method one gets the new variable which is
simply the linear combination of original variables that provides the
optimal separation in one-dimensional case. To characterize
discrimination capability of different parameters combinations after
application of Fisher's technique in Table~\ref{tab:fisher} we give
for them merit factors MF, areas $A$, separations $\langle S^2\rangle$
and misclassification rates $\xi$. Taking protons as `signal' and iron
nuclei as `background', one can consider $A$ as area under signal
efficiency versus background rejection curve~\cite{TMVA} (called also
receiver operating characteristics curve~\cite{webb}), the closer this
area to unity, the better the classification is. Separation is defined
in~\cite{TMVA} as
$$
\langle S^2\rangle=\frac12\int\frac{(\hat y_S(y)-\hat y_B(y))^2}{\hat y_S(y)+\hat y_B(y)}dy,
$$ where $\hat y_S(y)$ and $\hat y_B(y)$ are the probability density
functions for signal and background, $\langle S^2\rangle=1$ again
means the best separation and corresponds to distributions without
overlap. The misclassification rate, used in addition to these
statistical variables, is calculated in a very simple way to estimate
possible error in event-by-event classification approach. We fit
overlapping sides of distributions on Fisher variables for protons and
irons with Gaussian functions and consider all events to the left of
intersection point of these Gaussian fits as iron and all other events
as protons. In this case some number $\xi_\mathrm{p}$ of proton events
is recognized as irons and, vice versa, some number $\xi_\mathrm{Fe}$
of irons is classified as protons.

From Table~\ref{tab:fisher} one can see, that discrimination for both
high energy interaction models is very similar, though in case of
QGSJET~II the separation of p and Fe is slightly better, especially
for combinations of total signal and \xmax\ with LDF slope
parameter. Certainly, combination of \xmax\ with muon signal at 1000
meters provides the best discrimination, but as one can see
combinations of depth of shower maximum with the total signal in the
tanks in the range 700--1500 meters also provide excellent separation
of primaries with misclassification of only $\sim30-50$ events out of
2000. Further addition to this couple of other shower characteristics
does not improve significantly the discrimination capability and in
the case of the real data can be completely useless due to presence of
additional systematic errors, though the situation can change with
energy and zenith angle, of course. Taking into account robustness of
total signal at 1000~m to LDF reconstruction uncertainties the
combination (\xmax,\,\smilla) in our view looks as the optimal choice
for primary mass composition analysis in Auger experimental
conditions. Table~\ref{tab:fisher} also shows, that despite of week
discrimination power of charged particles density, its use together
with \xmax\ allows to achieve separation of primaries with MF=1.44
(for QGSJET~01), while for \xmax\ distributions alone merit factor is
equal to 1.16. At the considered energy and zenith angle
($D_{1000}$,\,\xmax) pair looks like the best choice for primary mass
reconstruction with TA.

Certainly, our conclusions on (\smilla,\,\xmax) and
($D_{1000}$,\,\xmax) as the best mass discrimination combinations are
specific only for the energy and zenith angle discussed, in the sense
that for another energies/angles addition of other parameters to
these basic pairs may be helpful in optimization of their mass
discrimination performance.
%\clearpage

\section{Extraction of composition from test samples with
  Multiparametric Topological Analysis}
\label{sec:mta}

\begin{figure}
\centering\includegraphics[width=\gsize\textwidth]{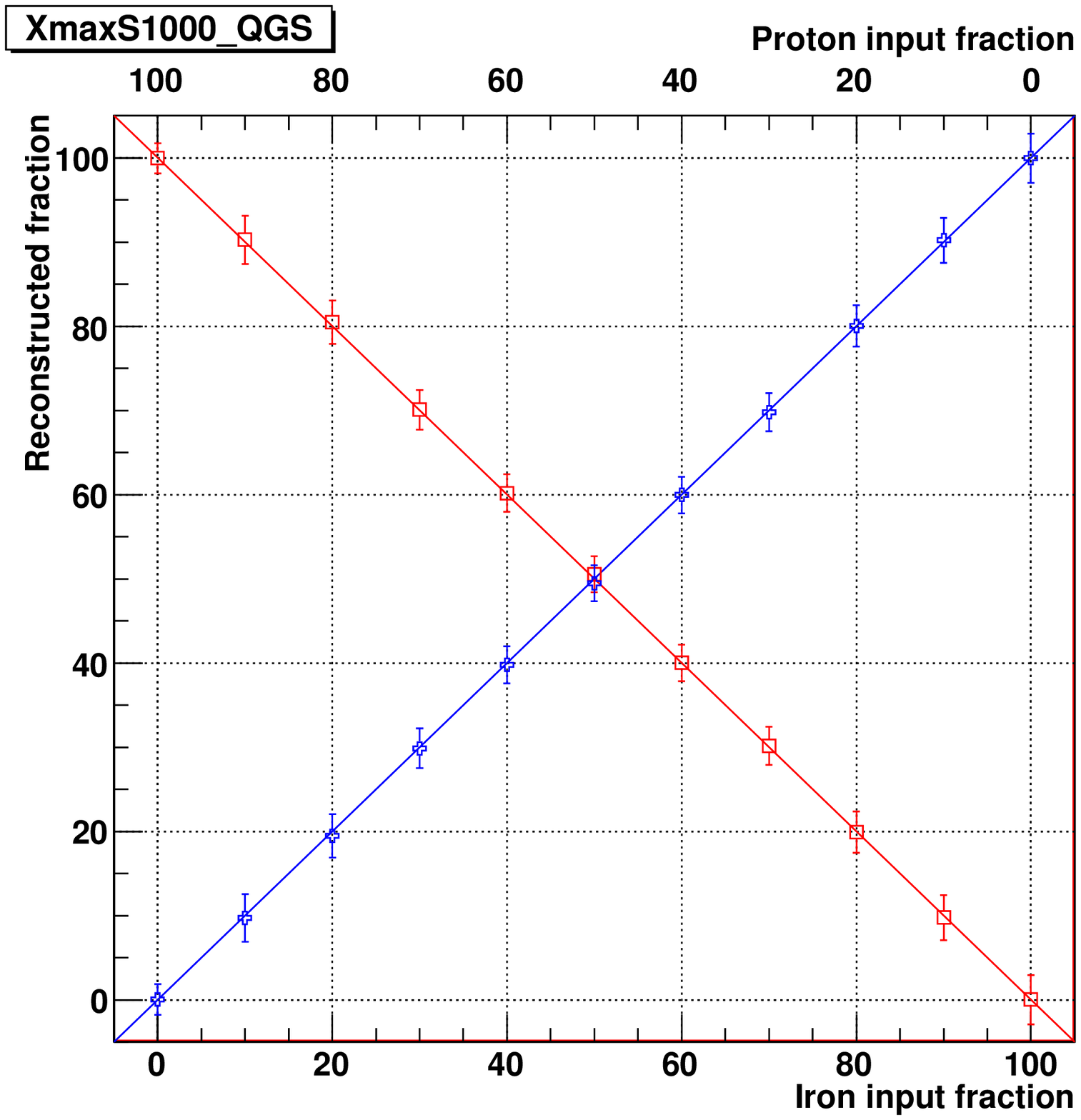}
\centering\includegraphics[width=\gsize\textwidth]{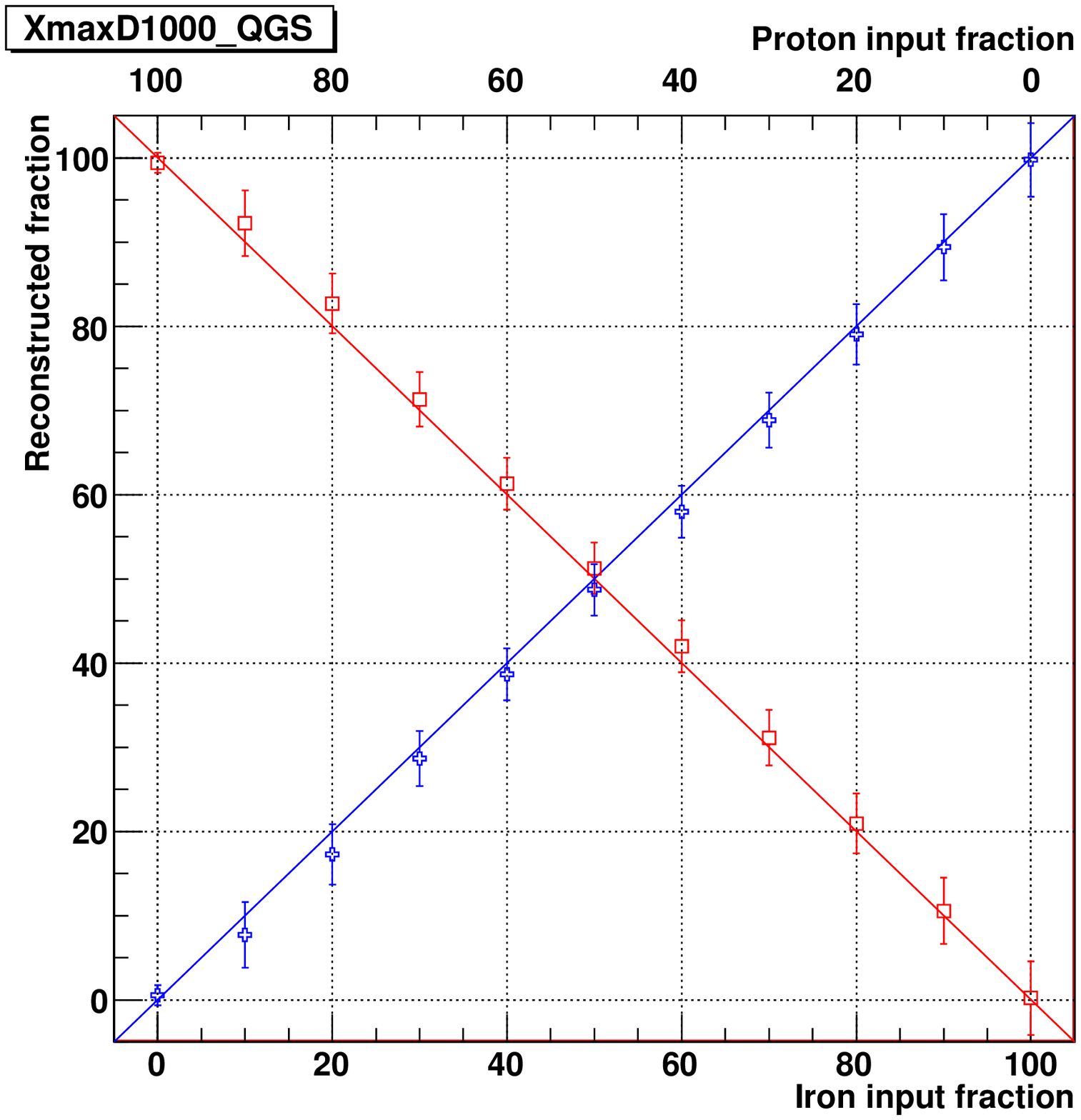}
\caption{Reconstructed with MTA on the basis of Fisher's variables
  F(\xmax,\,\smilla) and F(\xmax,\,$D_{1000}$) distributions proton (red
  squares) and iron (blue crosses) abundances in the samples with known
  primaries content. Lines mark the exact reconstruction results.}
\label{fig:mta}
\end{figure}

\begin{figure}
\centering\includegraphics[width=\gsize\textwidth]{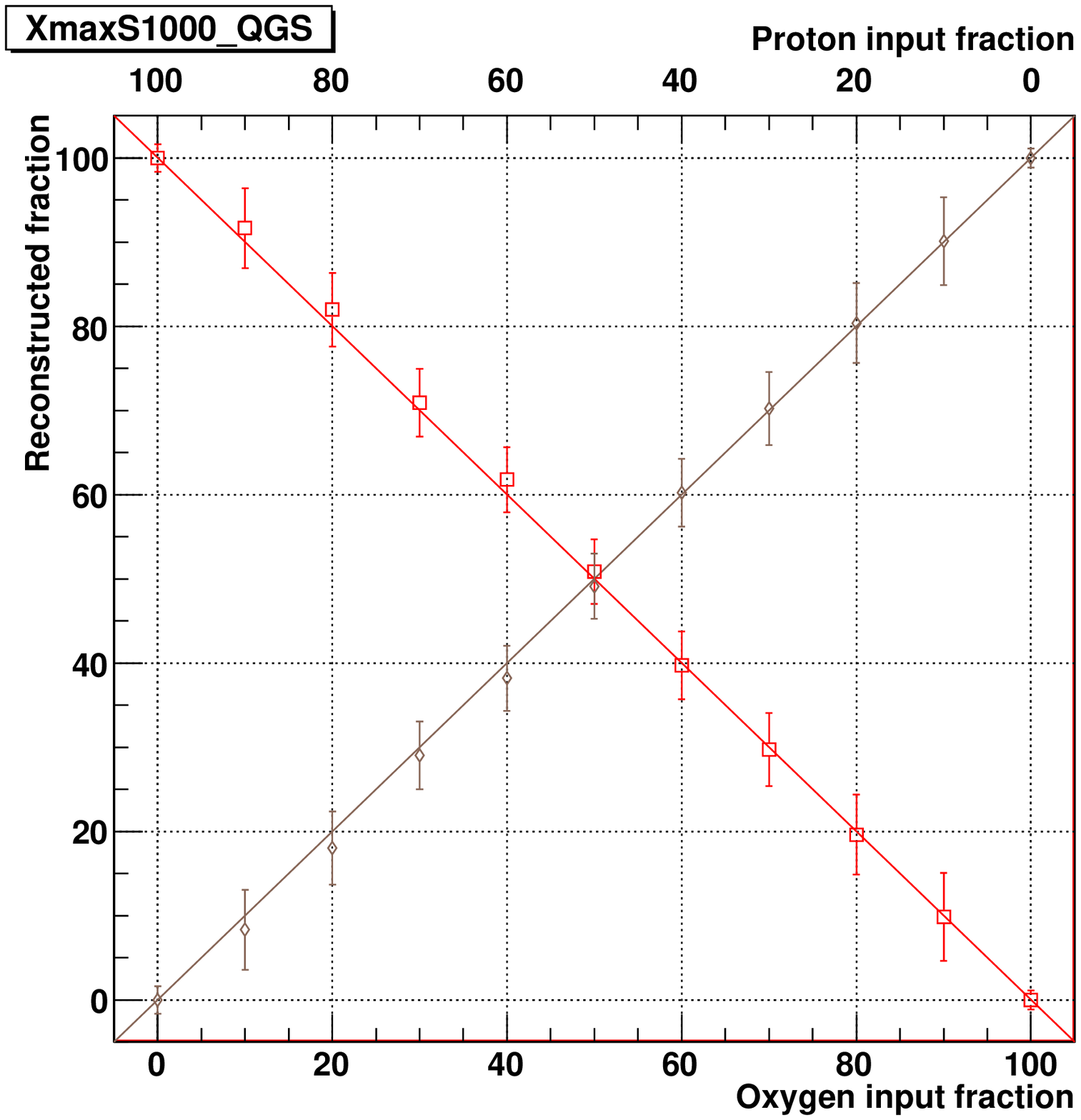}
\centering\includegraphics[width=\gsize\textwidth]{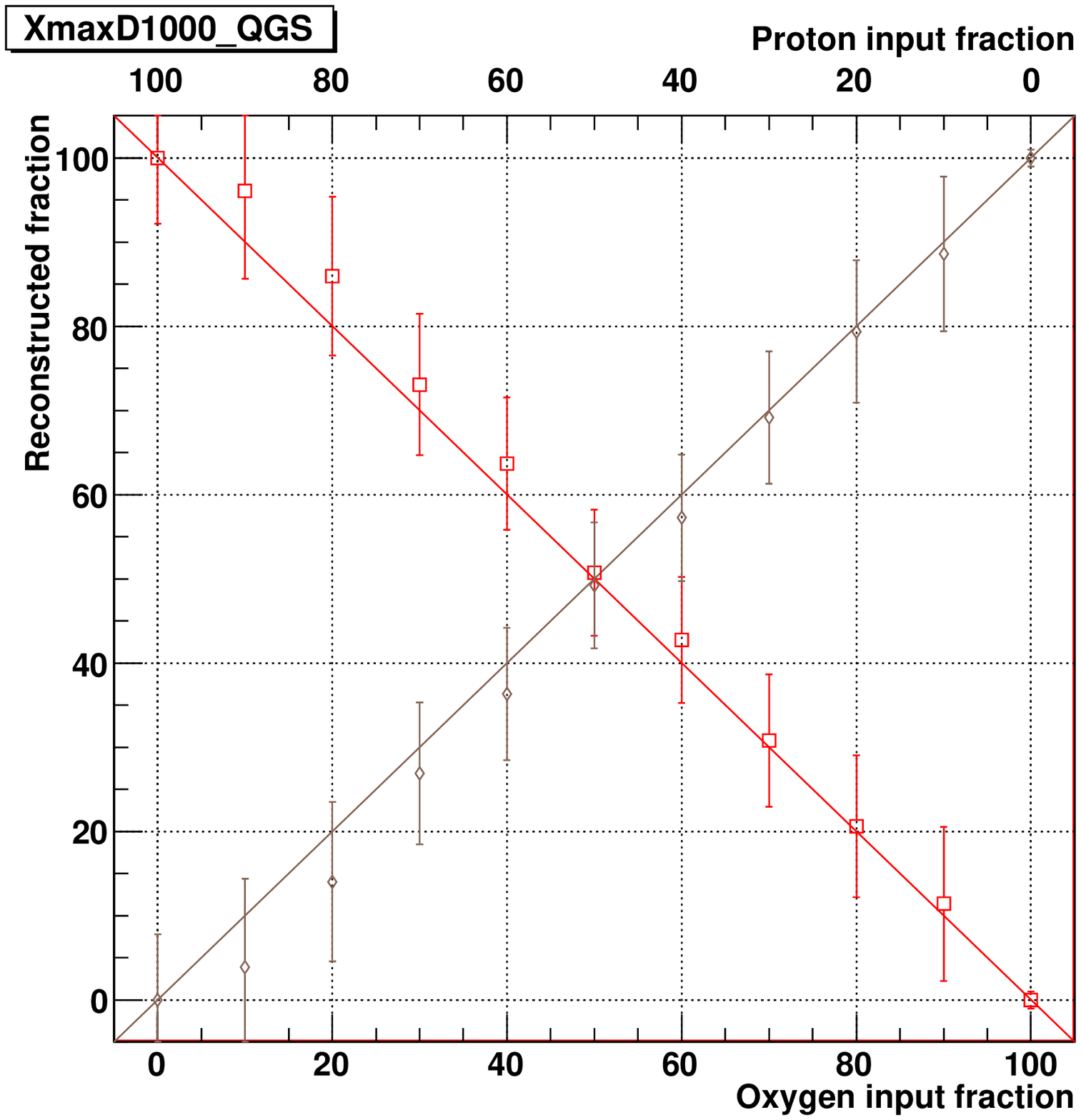}
\caption{Same as in Fig.~\ref{fig:mta}, but for proton (red
  squares)~--~oxygen (brown diamonds) mixtures.}
\label{fig:mtao}
\end{figure}

The basic idea behind the Multiparametric Topological Analysis
(MTA)~\cite{ambrosio_mta_2005} resides in the classification of
showers from different primaries according to their topological
distribution in multiparametric space. Considering
Fig.~\ref{fig:sigxmax} one can divide the plane (\xmax,\,\smilla) in a
number of cells and find probabilities for the showers falling in some
particular cell to be initiated by proton or iron. Using only these
probabilities on the pure set of proton showers one will erroneously
arrive (in case of the overlap of p and Fe populations) to mixed
composition. To correct such misclassification it is also necessary to
compute mixing probabilities~\cite{ambrosio_mta_2005}, determining the
chance of event from one primary mass in the given cell to be
misclassified as event of another primary mass. Hence, to get both
types of probabilities one has to use two independent sets of
simulated events. To illustrate classification capability of MTA
combined with discrimination power of Fisher's method, we have
performed primary composition reconstruction of sample mixtures with
known fractions of protons, oxygen and iron nuclei. In
Figs.~\ref{fig:mta},\,\ref{fig:mtao} we present the results of MTA
application to one-dimensional distributions on Fisher's variables
F(\xmax,\,\smilla) and F(\xmax,\,$D_{1000}$). The composition is very
well reproduced when one uses (\xmax,\,\smilla) combination, with
errors of 2--3\% for [p,\,Fe] and and 3--5\% for [p,\,O] mixtures. The
discrimination power of ($D_{1000}$,\,\xmax) couple is surely worse
(errors are 3--5\% for [p,\,Fe] and and 8--10\% for [p,\,O] mixtures)
and in case of real experimental conditions with additional systematic
errors its primary mass classification performance can be of limited
use in the case when [p,\,O] mixture is considered.

\begin{table}
\caption{Discrimination performance of different shower parameters
  combinations after application of Fisher's discriminant analysis.}
\label{tab:fisher}
\renewcommand\tabcolsep{5pt}
\renewcommand\arraystretch{1.2}
\begin{tabular}{|r|c|c|c|c|c|}
\multicolumn{6}{c}{QGSJET~01}\\
\hline
Parameters                                          & Area  & $\langle S^2\rangle$ & MF   & $\xi_\mathrm{p}$ & $\xi_\mathrm{Fe}$\\ 
\hline
[$S_{1000}^\mu$, \xmax]                                &  1.000&      0.995           & 2.53 &       10      &       2       \\

[$S_{700}^\mathrm{tot}$, \xmax]                         &  0.996&      0.908           & 1.90 &       35      &      15       \\ 

[$S_{1000}^\mathrm{tot}$, \xmax]                        &  0.996&      0.932           & 2.02 &        32     &       14       \\ 

[$S_{1500}^\mathrm{tot}$, \xmax]                        &  0.997&      0.940           & 2.18 &        33     &       14       \\ 

[$D_{1000}$, \xmax]                                   &  0.957&      0.677           &  1.44 &      139      &      65         \\ 

[LDF $\beta$, \xmax]                                &  0.925&      0.578           & 1.29 &       184     &       97       \\ 

[$S_{1000}^\mathrm{tot}$, LDF $\beta$]                  &  0.934&      0.627           &  1.49 &      172     &       78        \\ 

[\xmax, $S_{1000}^\mathrm{tot}$, LDF $\beta$]           &  0.997&      0.956           &  2.08 &       20     &        7         \\ 

[\xmax, $S_{1000}^\mathrm{tot}$, $N_\mathrm{max}$]        &  0.999&      0.946           &  2.16 &       25      &       11         \\ 
\hline
\end{tabular}

\begin{tabular}{|r|c|c|c|c|c|}
\multicolumn{6}{c}{QGSJET~II}\\
\hline
Parameters                                          & Area  & $\langle S^2\rangle$ & MF   & $\xi_\mathrm{p}$ & $\xi_\mathrm{Fe}$\\ 
\hline
[$S_{1000}^\mu$, \xmax]                                &  1.000&      0.985           & 2.70 &       11      &       1       \\ 

[$S_{700}^\mathrm{tot}$, \xmax]                         &  0.999&      0.961           & 2.18 &       24      &       8       \\ 

[$S_{1000}^\mathrm{tot}$, \xmax]                        &  0.996&      0.942           & 2.25 &        28     &       7       \\ 

[$S_{1500}^\mathrm{tot}$, \xmax]                        &  0.994&      0.937           & 2.32 &        21     &       11       \\ 

[$D_{1000}$, \xmax]                                   &  0.975&      0.770           &  1.65 &      99      &      54       \\ 

[LDF $\beta$, \xmax]                                &  0.947&      0.674           & 1.51 &       135     &       75       \\ 

[$S_{1000}^\mathrm{tot}$, LDF $\beta$]                  &  0.952&      0.718           &  1.64 &      124     &       73        \\ 

[\xmax, $S_{1000}^\mathrm{tot}$, LDF $\beta$]           &  0.999&      0.966           &  2.36 &       17     &        4         \\ 

[\xmax, $S_{1000}^\mathrm{tot}$, $N_\mathrm{max}$]        &  0.997&      0.953           &  2.33 &       23      &       5         \\ 
\hline
\end{tabular}
\end{table}

\section{Conclusions}
The present study allows to develop a new approach to the mass
composition analysis of hybrid data. We propose to use combinations of
longitudinal and lateral parameters to achieve maximum primaries
separation in multiparametric space. Further application of Fisher's
method optimizes discrimination, reducing the problem to
one-dimensional case and allowing for lower simulation statistics. At
the last stage one can apply different algorithms to extract mass
composition from distributions on Fisher variables, which are more
mass sensitive in comparison with e.g. traditionally used
\xmax\ alone. We applied for this purpose MTA technique to the samples
with different primaries fractions and retrieved with very good
accuracy nuclei abundances from [p,\,Fe] and [p,\,O] mixtures.

Regarding the choice of mass sensitive parameters, it was shown, that
for Auger the best mass discrimination can be achieved if to use
(\xmax,\,\smilla) pair, provided the primary energy is estimated from
the longitudinal shower profile. The charged particles density
measured in TA in combination with depth of shower maximum also
provides good discrimination of proton and iron showers, though in the
real experimental conditions its sensitivity seems to be limited for
proton-oxygen mixture case.

\section*{Acknowledgments}
We are very grateful to Maximo Ave and Fabian Schmidt for kind
permission to use their GEANT~4 lookup tables in our
calculations of signal from different particles in Auger water tanks.

%\bibliographystyle{/home/yushkov/bin/bibtex/IEEEbib} 
%\bibliography{/home/yushkov/bin/bibtex/my_rus}

\end{document}